\documentclass[letterpaper,twocolumn,10pt]{article}
\usepackage{main}
\usepackage{multirow}

\usepackage{tikz}
\usepackage{amsmath}

\usepackage{filecontents}
\usepackage{cite}

\begin{document}

\date{}

\title{\Large \textbf DangKiller: Eliminating Dangling Pointers Efficiently via Implicit Identifier}

\author{
 {\rm Daliang Xu}\\
 Peking University\\
 xudaliang@pku.edu.cn
\and
 {\rm Dongwei Chen}\\
Peking University\\
chendongwei@pku.edu.cn
\and
{\rm Chun Yang}\\
Peking University\\
yangchun@mprc.pku.edu.cn
\and
{\rm Kang Sun}\\
Peking University\\
ajksunkang@pku.edu.cn
\and
{\rm Xu Cheng}\\
Peking University\\
chengxu@mprc.pku.edu.cn
\and
{\rm Dong Tong}\\
Peking University\\
tongdong@pku.edu.cn
} 

\maketitle

\begin{abstract}
Use-After-Free vulnerabilities, allowing the attacker to access unintended memory via dangling pointers, are more  threatening. However, most state-of-art detection schemes can only detect dangling pointers and invalid them during the development and test phase, but not provide a tolerance mechanism to repair the errors at runtime. Also, these techniques obtain and manage the metadata inefficiently with complex structures and too much scan (sweep). The goal of this paper is to eliminate dangling pointers automatically and efficiently using compiler instrumentation with acceptable performance and memory overhead.

In this paper, we observe that most techniques lack accurate low-performance pointer graph metadata maintaining methods, so they need to scan the log to reduce the redundancy and sweep the whole address space to find dangling pointers. Also, they lack a direct, efficiently obtaining pointer graph metadata approach. The key insight of this paper is that a unique identifier can be used as a key to a hash or direct-map algorithm. Thus, this paper maintains the same implicit identifier with each memory object and its corresponding referent, associating the unique ID with metadata for memory objects, to provide schemes of obtaining and managing the pointer graph metadata efficiently. Therefore, with the delayed free technique adopted into C/C++, we present the DangKiller as a novel and lightweight dangling pointer elimination solution. We first demonstrate the MinFat Pointer, which can calculate unique implicit ID for each object and pointer quickly, and a shared-hash algorithm to obtain metadata. Secondly, we propose the Log Cache and Log Compression mechanism based on the ID to decrease the redundancy of dangling pointer candidates. Coupled with the Address Tagging architecture on an ARM64 system, our experiments show that the DangKiller can eliminate use-after-free vulnerabilities and provide full temporal memory safety at only 11\% and 3\% runtime overheads for the SPEC CPU2006 and 2017 benchmarks respectively, except for unique cases.
\end{abstract}

\section{Introduction}
Use-after-free (UAF) is the leading cause of memory temporal errors, which means a program uses pointers pointing to deleted objects, i.e., dangling pointers. According to Microsoft's annual vulnerability report, UAF is among the top 3 root causes of recent attacks \cite{RN171}. And it is a typical and common one exploited for return-oriented programming \cite{ RN169,RN33, RN38, RN170} and, data-only attacks \cite{RN174}.

Over the past decades, researchers have proposed many mitigation mechanisms against UAF vulnerabilities. Most of them are detection techniques with unbearable performance and memory overhead. Once applying such techniques to online services, services will be slower or even unavailable. In addition, they also require programmers to modify and refine the source code during the development and maintenance phases. However, programs today are more and more complex with their functionality increasing, which in turn increases the difficulty of fixing errors \cite{RN177}. What's more, due to the difficulty of finding dangling pointers, it is more tough to repair the bugs than the buffer overflow and uninitialized memory reading \cite{RN23,RN180,RN28}, so even if a vulnerability is detected, these vulnerabilities cannot be easily resolved. While eliminating techniques using delayed free, such as garbage collection, can eliminate the possibility of terminating the execution because of memory errors  \cite{RN54}. So, the service will not crash but may have some unanticipated result of one request. Thus, an error in the computation for one request tends to have little or no effect on the computation for subsequent requests \cite{RN176}.  So the online service can be more reliable. Therefore, this paper's goal is to provide a novel and lightweight dangling pointer elimination solution.

One of the mainstream elimination techniques \cite{RN11}, periodically scans the entire address space to find a dangling pointer. Since it is a sweep garbage collector without a pointer graph, it needs to sweep the whole memory area to confirm whether some memory locations store the reference to the object to be freed \cite{RN11}. To reduce the performance overhead, it uses multiple threads periodically synchronously to sweep the address space, and may result in memory leak. And there are some methods follow its sweeping idea \cite{RN178,RN179}, but no matter how they reduce the overhead, they still cannot avoid scanning the whole memory area to find dangling pointers. However, on resource-constrained embedded devices or high-loaded server devices, multi-threading techniques can put more pressure on devices, especially memory reads and writes. A pointer graph can keep track of all the pointers to each memory object so that the location of the potential dangling pointers can be detected accurately without scanning. Therefore, this paper hopes to combine the pointer graph with the elimination techniques.

\begin{figure}[ht] 
\begin{center}
  \includegraphics[width=0.48\textwidth]{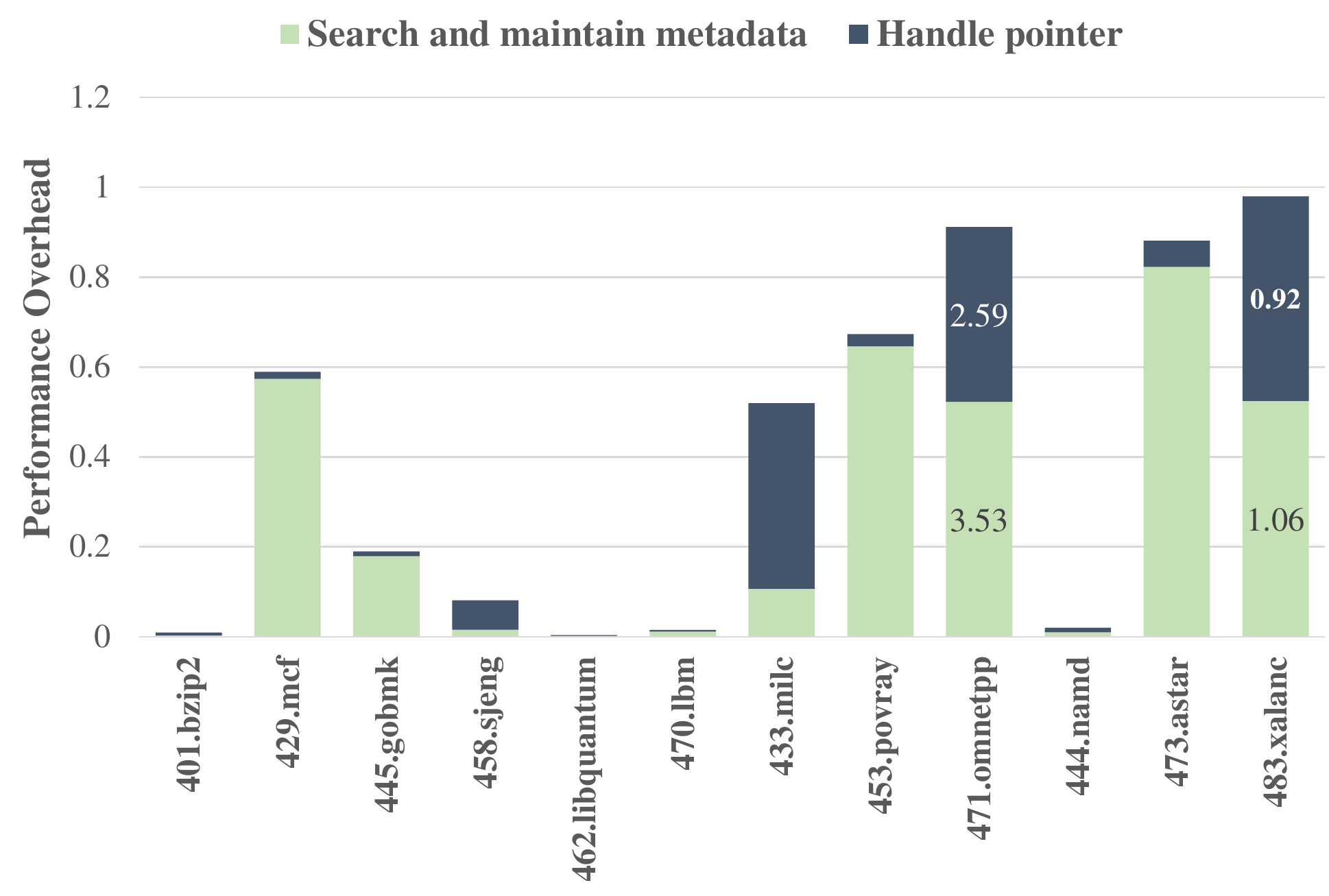}
  \caption[]{DangSan Performance Overhead On SPEC CPU2006}
\end{center}
\end{figure}

Various detection schemes shield software from dangling pointer exploitation \cite{RN23, RN54, RN52,RN28} by using pointer graph. Using logs of pointer locations linked to each object or reference counts, these approaches can obtain the locations of the potential dangling pointers in the memory. We found that the general process of pointer tracking could be divided into three steps: (1) Search for the corresponding object log metadata when a heap pointer is to be stored in memory; (2) Search the log and update the metadata if the memory location has not been recorded. (3) When freeing an object, if there are dangling pointers, use different approaches to handle those dangling pointers. However, these works tend to incur an excessively high performance overhead which is attributed to the inefficient search and maintenance of the pointer graph. We evaluated Dangsan, one of the most efficient state-of-art work, as shown in Figure 1. We find most benchmarks with higher performance overhead have occupied more than half overhead for search and maintenance, and some can even reach more than 90\%.  The main reasons for the high computational costs are as follows: (1) Lack of a direct mapping between objects metadata and pointers referring them. Dangsan \cite{RN52} relies on complex shadow memory structures to find metadata, while CRCount \cite{RN54} uses the interface from the memory allocator. Both of them incur unacceptable performance overhead. (2) Vast redundancy in the number of the candidate dangling pointer locations. To reduce the redundancy, these methods often scan lookup table in the log, leading to tremendous cost.

Our insight is that these problems can be solved by letting pointers and their intended referent share the same unique implicit ID. With the help of such IDs, we could quickly obtain the intended referent of a pointer, and immediately reach the metadata indicating an object's lifetime with hash or direct-map algorithm. As for the vast redundancy and reducing redundancy procedure, if we have the unique ID, we could build a cache to determine directly whether a particular memory location has been recorded in the log by combining the memory location with its ID. What's more, the implicit ID should be more regular for hash algorithm and not occupy independent memory area, compared with CETS \cite{RN24} lock and key.
\begin{figure*}[ht] 
\begin{center}
  \includegraphics[width=\textwidth]{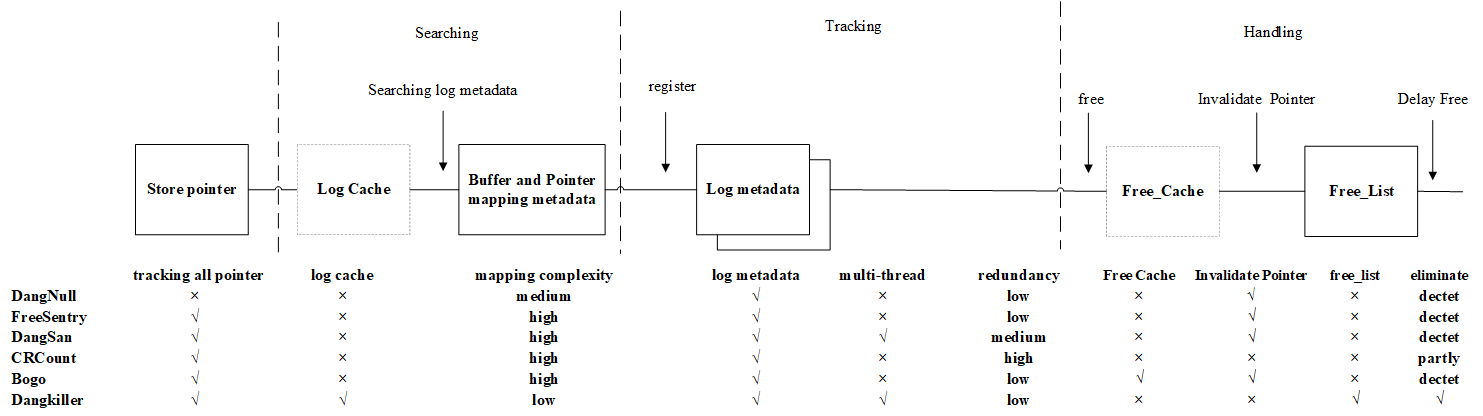}
  \caption[{aboveskip=0.5em, belowskip=0em}]{Comparison of temporal memory safety solutions}
\end{center}
\end{figure*} 
In this paper, we propose the DangKiller, which is a UAF vulnerabilities elimination system that is applicable to efficiently protect real-world C/C++ programs without modifying source code. Our elimination system combines the fast and efficient pointer tracking model with the delayed free technique to prevent the generation of dangling pointers during program execution with an acceptable performance and memory overhead. The overhead of DangKiller is lower than most of the existing detection and elimination techniques.
 
To give the memory object and its corresponding pointers a unique implicit ID, we demonstrate the MinFat Pointer \cite{RN163,RN8,RN183,RN184}, which is a simple Tagged Pointer scheme. MinFat Pointer constrains the sizes of allocated memory and their alignments to powers of two, and store $log_{2}(size)$ in the high bits of the pointer. With such pointers, the Dangkiller can obtain the base address of an object easily and use it as the implicit ID of the object and pointers, assuming no spatial memory errors. And a shared hash algorithm and log structure are also presented to obtain the object metadata based on the ID to help reduce the performance overhead associated with metadata searching. To reduce the location redundancy and complexity of scanning the lookup table, we propose the Log Cache. The Log Cache is a metadata structure to determine whether the memory location has been stored in the object log quickly by combining location with the ID. However, the tag on high bits also introduces compatibility problems, because a tagged pointer can not be used to access memory directly. These tags need to be masked off, which will bring significant performance overheads. Luckily, this problem has been solved by the Address Tagging proposed by ARM/SPARC systems. In particular, ARM has added the Address Tagging to the ARMv8 ISA which is designed for servers and desktops \cite{RN167}. ARMv8 architecture ignores the highest 8 bits of a pointer when accessing memory. Some existing work, such as HWASAN \cite{RN166}, uses the same technique to reduce the performance overhead and improve its compatibility. We adopted the delayed free technique, and optimized the free\_list structure. Objects are freed until all the references of the objects disappear by themselves during program execution.

We make the following contributions:
\begin{itemize}
\item We propose DangKiller, the first dangling pointer elimination technique combined with Pointer Graph in C/C++. Our system can run with a reasonable performance overhead and has good scalability.
\item We propose a scheme to correlate pointers with their intended referent by having them to share the same implicit unique ID. A tagged pointer system named MinFat Pointer is integrated to support the quick calculation of IDs. Furthermore, we utilize the Address Tagging feature in ARM64 architectures to relieve the compatibility problems and performance overhead caused by tagged pointers.
\item We propose a shared hash algorithm and log structure to reduce the overhead of searching and maintaining the pointer graph. We also provide pointer Log Cache, a metadata structure verifying redundancy quickly. Finally, we propose Log Compression to reduce the redundant reference further.
\item We use LLVM-LTO pass on the llvm-3.8 platform and combine it with a static analysis method to achieve the goal of eliminating the dangling pointer. We evaluate DangKiller and show it is able to eliminate dangling pointers with acceptable overhead and even lower than state-of-the-art work.
\end{itemize}

\section{Background and Related Work}
Most of the dangling pointer detection tools are used during software development. After discovering errors causing the program to crash, programmers have to patch the source code, which will consume a lot of resources, seriously damaging the productivity of software development. On the other hand, the elimination technique can tolerate dangling pointers by leaving an object undeleted until all the pointers referring it vanish by themselves during the program execution, i.e. delayed free.

Temporal memory safety violations occur when a program accesses a deallocated object. There has been several solutions proposed for temporal memory safety. Memory allocator \cite{RN9, RN15,RN25} attempts to prevent allocated objects from ending up at the same address as previously freed objects. \cite{RN7, RN128,RN36, RN37} used pool allocation for the same purpose. Other works \cite{RN8,RN26, RN49} track per-object metadata: e.g., ASAN \cite{RN26} poisons deallocated objects, and report access to the poisoned area. Also, some static analysis approaches \cite{RN16} were proposed, but they could only recognize relatively simple cases and were therefore inherently prone to false negatives. 

\subsection{Pointer graph based scheme}
Most schemes use the pointer graph-based method \cite{RN23, RN54, RN52,RN28}. They keep track of the pointers for each memory object and build a pointer graph where the nodes and edges are the objects and their connections, respectively. Figure 2 shows such a method, which generally has the following three steps. (1) Searching: when a pointer is about to be stored in the memory, the pointer is used to find the memory object pointed to and obtain the corresponding pointer graph. (2) Tracking: update the reference in the pointer graph based on the memory location of the pointer. (3) Handling: when the memory object is freed, the reference retained in the pointer graph is used to prevent the dangling pointer from being accessed.  
\newline
\newline\textbf{Searching.} Mappings between objects and their metadata have been widely proposed. The DangNull \cite{RN23} uses a thread-safe shadowed red-black tree to locate the metadata corresponding to the memory object with an unacceptable performance overhead when the entire memory area is tracked. The FreeSentry \cite{RN28} is based on the label lookup table and continuously finds log metadata on multiple levels. The DangSan \cite{RN52} uses two levels of the shadow memory\cite{RN19} to acquire the log metadata. Similarly, Bogo \cite{RN53} uses MPX's two-level Bound Tables \cite{RN44}, which are both complicated. Some other works \cite{RN54} rely on the interface provided by the memory allocator to obtain the object identity and locate the metadata. Regardless of the complex data structure and the interface provided by the memory allocator, the pointer to object mapping is complex, and has a large performance overhead. This is mainly because the priories do not have an efficient method of correlating a pointer with its intended referent and do not share a same ID for each memory object. 
\newline
\newline\textbf{Tracking.} There are many approaches proposed to maintain or update the  pointer graph in different ways. The FreeSentry and DangNull transfer the original references to the new object based on its location. To reduce the overhead of registering reference, CRCount \cite{RN54} only retains the reference count without maintaining the reference graph. However, the reference count alone cannot adequately describe the reference between the memory objects. While DangSan only registers new references but do not remove the original one. To reduce the redundant references to the same object, this approach uses a lookup table and hash table to verify whether the reference has been recorded. As a result, the performance overhead is significant. Therefore, this paper hopes to propose a pointer graph with a low maintenance overhead and minimal redundancy.
\newline
\newline\textbf{Handling.} When the user frees an object, there are two main schemes that can be adopted. Most detection techniques nullify all the pointers pointing to the object being deallocated. When a dangling pointer is dereferenced, a segmentation fault is triggered to prevent the program from continuing to run. In contrast, elimination methods usually delay the free of objects until there is no reference to the object. CRCount uses delayed free, which frees the object gradually. Since the reference count can lead to deadlock, CRCount can only partially eliminates the dangling pointers. In this paper, we hope to propose a technique that eliminates all the dangling pointers using the delayed free technique.

\subsection{Identifier based scheme}
There are also identifier-based schemes \cite{RN43, RN24} using a lock and key to identify pointers and objects. This type of method allocates an explicit, non-repeating ID for each object and stores it in separate metadata. This method allows the program to determine quickly whether the object and pointer are corresponding to each other. Thus, the load/store operations is only allowed when the lock and key are the same. Although this method can quickly detect objects and pointers that do not correspond, obtaining the explicit ID has a relatively large performance overhead since IDs are stored in a disjoint metadata structure. Meanwhile, the separate metadata storage ID will also bring a certain memory overhead.


\subsection{Address Tagging}

A tagged pointer is often subjected to problems due to its inability to access memory directly. When a Tagged Pointer is used to access memory, the tag needs to first be masked. Address Tagging provides architecture level support for Tagged Pointer \cite{RN167}. In particular, ARM added the Address Tagging to the ARMv8 ISA, which is also the default configuration option of the system. Using this technique, the processor can ignore the upper 8 bits of the pointer during memory access. Some existing works, such as HWASAN \cite{RN165, RN166}, use this technique to store the tag that corresponds to the object in the upper 8 bits, which reduces the memory access overhead and improves the compatibility. Thus, once Address Tagging implemented, the Tagged Pointer can directly access the memory without masking.

\section{Motivation}
The goal of this paper is to eliminate dangling pointers with delayed free technique. To reduce the performance overhead, we generate a unique ID for every object, and build a pointer graph with high efficiency based on those IDs.

Figure 2 shows the general process of detecting and eliminating dangling pointers. Whether a pointer is a dangling pointer or not, it has to go through all three steps, as mentioned in \$2.1. Tracking can be further divided into two steps: verifying and logging. Verifying checks whether is reference is redundant. Logging inserts a reference into the log. Therefore, we can calculate the overhead with the following equation.N means the number, while T means the time overhead.
\newline

$T_{overhead} = N_{pointer}*(T_{searching}+T_{verifying}+T_{logging}+T_{handling})$ \quad (1)
\newline
\newline\textbf{Overhead Quantitative Approach.} The $N_{pointer}$ means the number of pointer after reducing the redundancy. For consistent cache hit rate and system access memory latency, the $T_{handling}$ and $T_{logging}$ are fixed during any program execution. Therefore, the key to get past the performance bottleneck is the $N_{pointer}$, $T_{searching}$ and $T_{verifying}$, which are related with the efficiency of metadata search and reference verification. We analyze these two part: (1) reducing the redundancy leads to the increase of $T_{verifying}$. For example, Dangsan uses lookup table to reduce the redundancy of references. Increasing the number of lookup table will bring lower redundancy, and increase $T_{verifying}$ at the same time because Dangsan needs to go through all the tables. To solve this problem, we use a direct-map cache based on implicit ID to replace the lookup tables used by most approaches. And our Log Cache can provide lower redundancy rate and the less $T_{verifying}$.  (2) to reduce $T_{searching}$, we also provide a easier searching methods by using hash and direct-map algorithm based on the implicit ID.
\newline

\textbf{Implicit ID.} To address the above problems, we propose the implicit ID. We can be sure that any heap object will have a base and bound address within the original allocation, because the bounds of an object can never be enlarged, only restricted. Thus, the base and bound address of the object can be a unique address for each different objects during the program execution. Since  base addresses often can be obtained more easily, our insight is that we use the base address of an object as its unique implicit ID to represent the object and all the subordinate pointers.

To get the base address of one object, many methods have been proposed, such as MPX \cite{RN44}, CHERI \cite{RN49}, and tagged pointer. Actually, any approach that can obtain the base address of an object is suitable for DangKiller. Therefore, our method can be greatly compatible with existing spatial memory detection schemes. However, if we only care about the temporal safety, those schemes have many drawbacks. MPX has to search the page-based boundary table, and CHERI changes all the hardware architecture to support the capability model. As for tagged pointer, usually the highest 16 bits of 64-bit address space are used to store metadata information \cite{ RN164,RN40, RN163}, which can be used to obtain the base and bound of the object. The LowFat Pointer \cite{RN163} needs to calculate the base and bound addresses according to the BIMA encoded metadata. While this saves a significant amount of memory, it increases the computational complexity and is not suitable for the existing memory free environment. The other works \cite{RN8, RN162, RN161} embed the tag into a valid address space to improve the compatibility and divide the program's virtual address space into several regions of equal size. The object size is then allocated to powers of two to improve the efficiency of calculating the base and bound. However, the implementation of the region is too complicated, and the memory allocator requires modification.
\newline

\begin{figure}[ht] 
\begin{center}
  \includegraphics[width=0.48\textwidth]{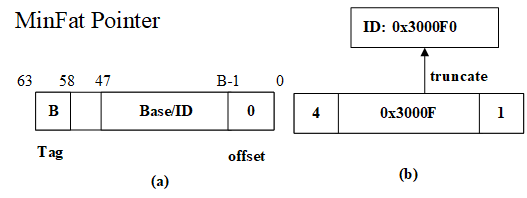}
  \caption[]{MinFat Pointer and its example}
\end{center}
\end{figure}

\begin{figure*}[ht] 
\begin{center}
  \includegraphics[width=\textwidth]{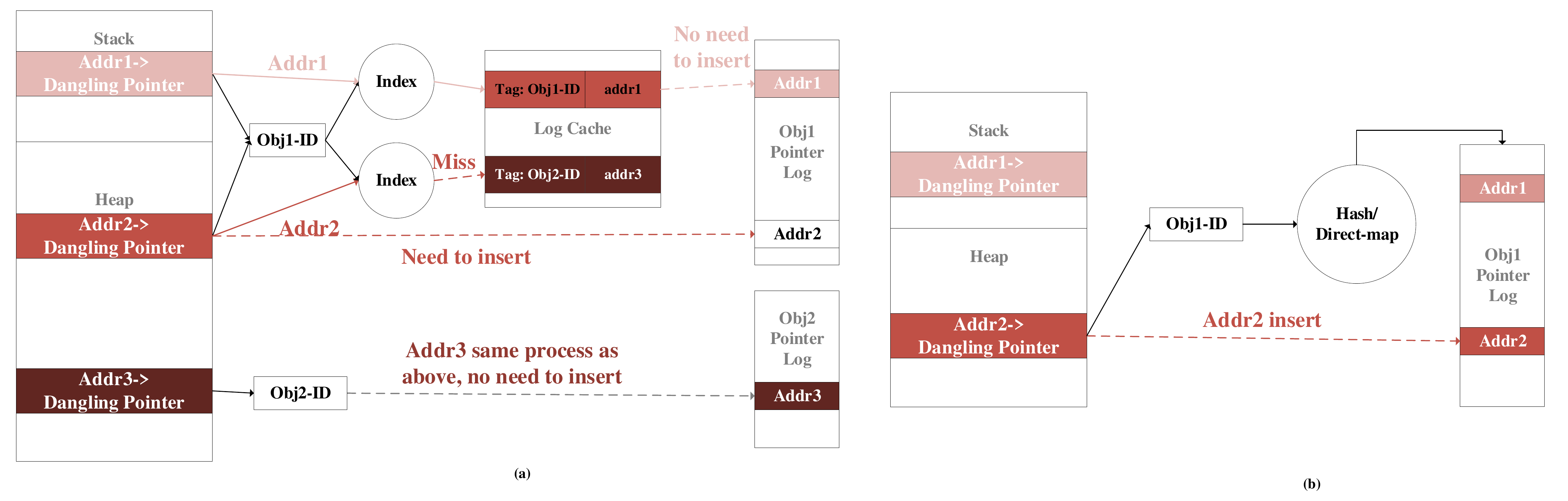}
  \caption[]{DangKiller component structures. (a) shows the process of verifying whether a dangling pointer is redundant. (b) shows the process of inserting a new reference into the log metadata.}
\end{center}
\end{figure*}

\textbf{MinFat Pointer.} To minimize the procedure, to make the ID data more regular, and to be suitable for resource-constrained scene, such as IoT devices, we demonstrate MinFat Poiner, as shown in Figure 3(a). It uses higher-order bits of a pointer to store size metadata so that the base can be calculated by truncation. Figure 3(b) shows two MinFat Pointer p. For example, the pointer p is 0x3000F1, with a tag 4 in it. The tag 4 means when calculating the ID, we should truncate the lowest 4 bits. So the ID 0x3000F0 is calculated. From the example, we can see that the ID can be quickly calculated just by arithmetic operation. Compared to the explicit ID, such as CETS, our implicit ID does not occupy any memory space and needs no sophisticated structure to obtain the ID. Once our implicit ID is calculated, we can use the hash and direct-map algorithms to obtain the metadata quickly. Also, it is possible to use ID and location to detect if a location has been stored before obtaining the metadata so that redundancy can be greatly reduced with lower performance overhead.The ARM64 architecture supports Address Tagging, which ignores the highest 8 bits of the 64-bit address during memory access. Therefore, MinFat Pointer only does little harm to compatibility and performance.

\section{Overview}
\subsection{Threat Model}
Firstly, we believe processors can be trusted, and there are no hardware security bugs in the processor circuit fabrication \cite{RN22, RN42}. We assume that the vulnerable program which only has some dangling pointers may contain one or more temporal vulnerabilities, such as use-after-free vulnerabilities where an attacker can read from or write to the memory area which has been freed to the memory management system and double\_frees or frees of memory, not on the Heap vulnerabilities, which can corrupt memory management data structures. To some other memory errors (e.g., buffer overflows), We assume that they are already defended by other Sanitizers such as ASAN \cite{RN51, RN35, RN37, RN26}. Expressly, we assume that the attacker cannot access or destroy our data structures, due to orthogonal defenses such as hardened information hiding, ASLR \cite{RN17, RN18, RN21}, SGX \cite{RN57}, or other efficient isolation techniques.
\subsection{Components of DangKiller}
DangKiller is primarily composed of two modules: the pointer tracking module, and the delayed free module. The pointer tracking module tracks the locations of dangling pointer candidates;and the delayed free module frees the object when it has no references. The modules are correspond in Figure 4 and Figure 5 respectively.

According to Figure 4, any dangling pointer will have an implicit obj-ID calculated by MinFat Pointer.As mentioned above, we divide the tracking procedure as three steps. Rather than searching, tracking, and handling in most of work, we follow the sequence of tracking, searching and handling. For tracking, whenever a dangling pointer is stored, by using its obj-ID and its location address, we can obtain an index of the Log Cache. If the tag is just correspond to the obj-ID, which means hit, we don't need to insert the location address in log metadata as red cache line and red dotted line shows. The addr1 has already stored in obj1 log metadata, so it don't need to store it again. And it is the same as the gray one. However, if the cache miss, we need to insert the addr2 in obj2 log metadata. So we need to step into searching phase. Before inserting the addr2, as the second picture shown in Figure 4, we first calculated its implicit obj-ID, and then using hash or direct-map algorithm to search for the log metadata. When the address of the metadata is available, we just insert the location address to the metadata in sequence.This part of the module relies primarily on the LTO (link-time optimization) pass of the LLVM \cite{RN175}.
\begin{figure}[ht]
  \includegraphics[width=0.48\textwidth]{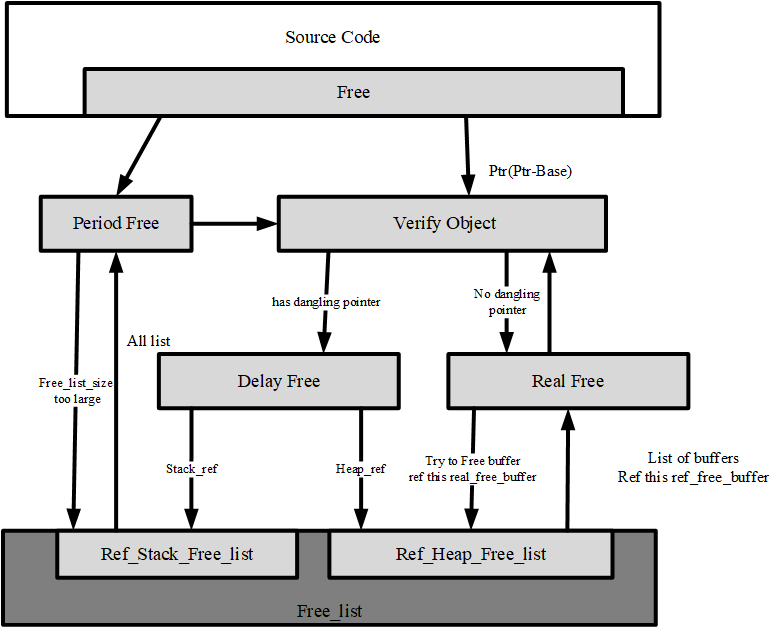}
  \caption[{aboveskip=0.5em, belowskip=0em}]{delayed free module overview.}
\end{figure}

From Figure 5, the process for to free a object, includes two parts: the right side displays a series of judgments on the object that the user wants to free, and the left side is the operation of delaying the free of specific objects in the free\_list periodically. The right side operation determines whether the user can release the object \emph{buf} using the module denoted as Verify Object. This module will verify whether there still some referring dangling pointers by using the log. If true, the object is released and the ref\_heap\_free\_list of the object is obtained. The list consists of many other objects that cannot be released due to the existence of the \emph{buf}, and then Dangkiller continues to verify the list. If the object cannot be released immediately, it needs to be released later. Then, relying on whether the location is in a heap, DangKiller selects the ref\_period\_free\_list or ref\_heap\_free\_list to record the \emph{buf\_id}. The Period Free, on the left side, isn't often executed, unless certain conditions met, see Section 4.6. This process is similar to the right side, except for the object from the ref\_period\_free\_list. This part of the module is developed using LLVM's compiler\_rt runtime library.

\section{Design}
\subsection{MinFat Pointer}
The MinFat Pointer is a tagged pointer that uses the highest 6 bits of a pointer to store the metadata information of the object size, as shown in Figure 3(a). The upper 6 bits store the upward rounded value of the logarithm for the size of the object corresponding to the pointer. That is $ B=\lceil \log_{2}size\rceil$.

To obtain the boundary of a memory object quickly, we enlarge the actual allocation size of the object to the nearest powers of two, denoted as AllocSize. To make this allocation possible and MinFat Pointer execution effectively, we designed MinFat pass and run-time-lib. To allocate a AllocSize object and return MinFat Pointer, we replace all allocation operations of the user program, such as malloc and new, with the MinFat\_Malloc function. This function, according to the AllocSize, allocate and align the AllocSize and return it to the user program with the tag. For the user's release operation, such as free or delete, DangKiller will replace it with MinFat\_Free,  and a series of delayed free operations will be performed before calling libc\_free. The realloc process may be relatively complicated because realloc means that a new object is allocated and the contents of the old object are copied into it. So to imitate the realloc of the libc, MinFat\_Realloc will call MinFat\_Malloc, memcpy, and MinFat\_free in sequence. If there is still a reference to the old object in the program, then the old object will not be released. Although it will bring some memory overhead, it is safer. Also, when the application is running, it will call a lot of std\-lib functions, such as open, printf, etc. Although standard library function is developed to be suitable for the ARM64 architecture, there is still potential poor compatibility, due to the tag in the pointer. Therefore, replace the std\-lib library function with the MinFat\_Wrapper  when called. The MinFat\_Wrapper will first mask the tag and then call the actual std\-lib function.

\subsection{Object-Metadata Address Table}
An essential part of the pointer tracking module is maintaining the mapping between memory objects and pointers to support a quick search of metadata. We want to use ID as an index to access the object-metadata address table and directly obtain the log metadata address corresponding to each object. Since the valid address of the pointer is 48 bits and the ID is also 48 bits, the ideal index has a length of 48 bits, so a table sized at $2^{48}*64$B is needed. However, such an implementation is impossible. To reduce the memory overhead and ensure that the performance overhead is acceptable, we use a combination of the direct mapped and hash techniques to reduce the address table size and minimize the performance overhead.

\begin{figure}[ht]
  \includegraphics[width=0.48\textwidth]{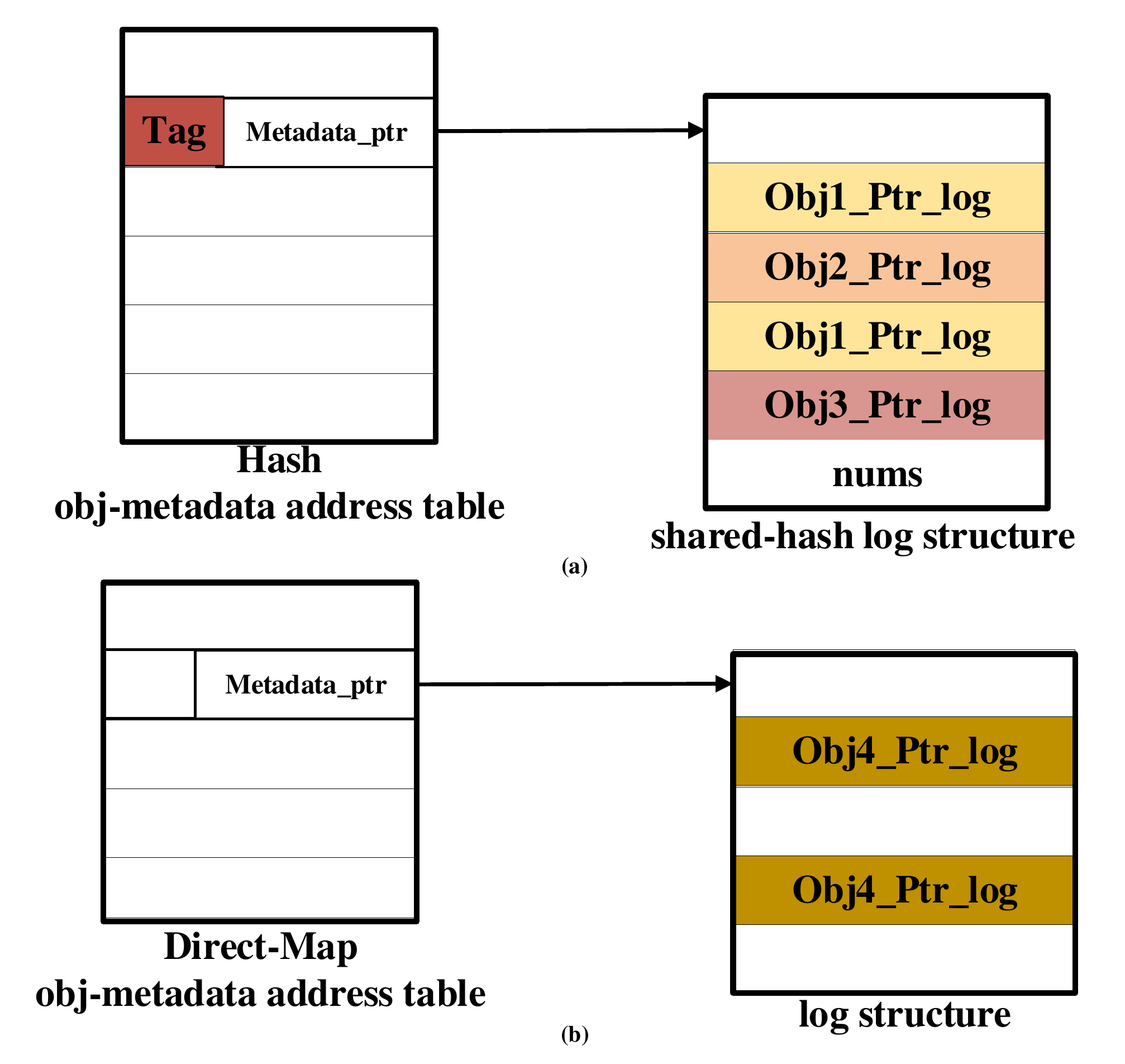}
  \caption[{aboveskip=0.5em, belowskip=0em}]{Object-metadata address table and log structure overview. (a) is in hash circumstance while (b) is direct-map.}
\end{figure}

The direct mapped method directly indexes the object-metadata address table using the ID. In this way, the overhead of the search metadata is relatively small. Therefore, we hope to organize as many objects as possible in this way. We investigated most of the memory allocators, such as the tcmalloc \cite{RN181} and ptmalloc \cite{RN182}, and found that nearly all of them use the lower address space for allocations. Thus, when a pointer corresponds to the ID within a $2^N$ (N is a threshold parameter to control shared-hash algorithm, and we will detail N in \$7) address space, we use the direct mapped method indexes the object-metadata address table to obtain the log metadata corresponding to the object. Whenever a object is allocated in the address space within $2^N$ by using Minfat\_Malloc, an entry in a table and the log metadata are assigned according to the ID; similarly, when the object is indeed released (not user called), it is also necessary to release the metadata. As shown in Figure 6(b), each table entry pointer to the address of a log structure which only contains one object metadata.

In some cases, because the object is allocated in a relatively higher address space, we can not use the direct mapped method. In this case, a hash algorithm is necessary, although this causes the problem of hash collisions. As multiple objects may share the same entry of an object-metadata address table, we record the extra sign in the entry to represent that the entry corresponds to an ID, and the next object will share that entry, As Figure 6(a) the red tag shown. Moreover, because the actual address space uses only 48 bits of the entire space, we use its high 16-bit or 8-bit storage the sign, a fixed random value generated by a program to reduce the memory overhead. Choosing 8-bit tags on the ARM64 platform allows for excellent compatibility. Although, the 8-bit tag may cause the same probability of the garbage value increasing in the memory area, so the tag is randomly generated and allocated, and a new page is filled with 0. Therefore, we believe an 8-bit tag is sufficient and the actual test does not appear to be equal.

At a high level, when an object is created, and the address obtained from the table contains a tag, the object is considered to have used the same entry, and the log-structured metadata needs to be shared. To ensure the correctness of the log metadata, when multiple objects share a log, if one of the objects is freed, the log metadata corresponding to the object will not be cleared. Only after all the objects corresponding to the entry are freed, will the space be released together. Determining when the log metadata will be released is explained in \$5.3.

\subsection{Shared-Hash Log Structure}
Now that we have a mapping relationship of the pointer-to-object, storing such metadata is especially important. This is particularly true if multiple objects share the same log metadata due to hash collisions. 

We use the shared-hash log structure to maintain the metadata in the scenario of hash collisions, as shouwn in Figure 6(a). The log structure is similar to the log structure proposed by DangSan. The main difference is that each log corresponds to a thread with an extra attribute \emph{nums}, which represents the number of objects share the log-structured. The \emph{nums} in other threads has no practical meaning. When an object is allocated, DangKiller first determines whether the mapping relationship of the shared ptr-to-obj is generated. If so, the first address of its shared-log metadata is obtained and then \emph{nums} is increased by one. If it is a new mapping, a new entry will be added to the object-metadata address table and assigned a shared log-structured metadata with \emph{nums = 1}. Only when all the objects corresponding to the log-structured metadata are freed \emph{nums} is  reset to zero; thus, the space is only released together. Also, the tag in the tagged pointer is removed from the object-metadata address table and the value stored in the entry is set to 0 to indicate that the entry is unoccupied. There is no hash collision and no sharing is required. When using multithreading, the \emph{nums} operation may involve atomic operations, but this only occurs when the object is released and allocated, indicating the performance overhead is relatively small.

It is noted that the same thread log metadata may store the location of the potential dangling pointers for different objects. As Figure 6(a) shows, the log is shared with objects 1, 2, and 3; object 1 has two potential locations, object 2 has one potential location, and object 3 has one potential location. Then, after object 1 is freed, the Verify Object process judges whether the dangling pointer is still stored in the location according to its base and bound addresses. Although multiple objects that correspond to a pointer share a log, the reference of object 1 is not stored in the memory location which the object2\_ptr\_log correspond to. Thus, although we seem to verify more potential memory locations, it won't make false positives. Also, in fact, experiments show that the number of hash collisions is relatively small, and there are usually only two objects that share one object-metadata, so log metadata will not harm the performance of searching and inserting procedure.
\begin{figure}[ht]
  \includegraphics[width=0.48\textwidth]{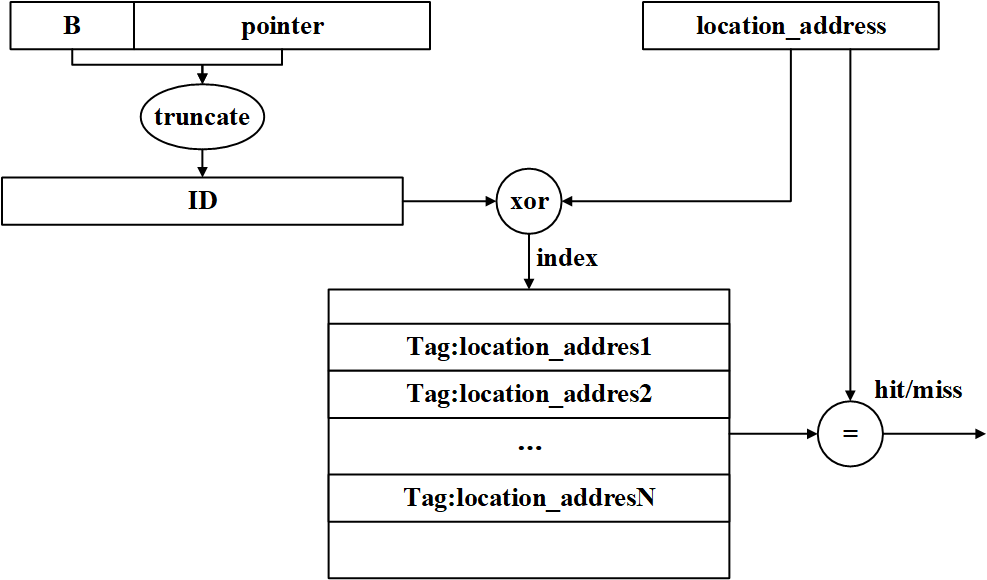}
  \caption[{aboveskip=0.5em, belowskip=0em}]{Log Cache structure}
\end{figure}
\begin{figure*}[ht]
  \includegraphics[width=\textwidth]{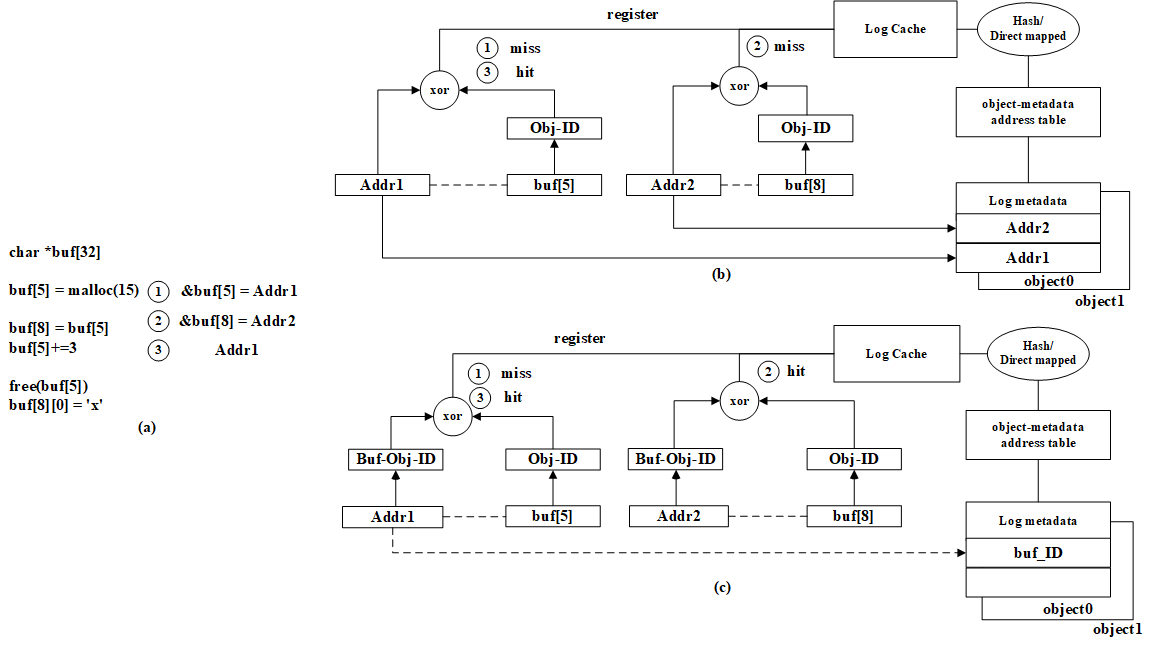}
  \caption[{aboveskip=0.5em, belowskip=0em}]{(a) shows a common example that dangling pointers shared same ID occurs in the same object many times, like buf[5] and buf[8]; (b) shows when Log Compression off; (c) shows when Log Compression on}
\end{figure*}
\subsection{Log Cache}
Another critical part of the pointer tracking module is reducing the location redundancy. The previous sections explained that in the pointer tracking process, there are several duplicate locations that may be stored, so we propose a Log Cache to reduce these duplicates before looking up the log metadata using the location and the object ID, as shown in Figure 7. The proposed Log Cache is an independent data structure, similar to the cache inside the processor. The critical problem is the indexing and replacement algorithm.

The Log Cache represents whether a location is in log metadata of an object. Thus, when accessing the cache, we care only about the object. Therefore, using the MinFat Pointer we can quickly obtain the object ID and index the cache.

Most of the internal caches of the processor use some bits in the middle of the pointer as the index and find the corresponding cache lines ,which is similar to DangSan's lookup table. However, unlike the hardware cache, which can access multiple lines at the same time, the Log Cache is a software cache and needs to traverse multiple lines using a loop, which results in a relatively large performance overhead. 

Besides,the ID and location are two factors that need to be considered. Thus, we need an index that contains both information pieces. Therefore, we design our Log Cache, and use the XOR value of the ID and location as an index. The relational algebra with the highest entropy, as shown in Figure 7, can simultaneously represent the object and location. Only the Tag is stored in each entry, which determines the location itself. When the cache is indexed, if the Tag is found to be at the same location, since the XOR is reproducible, we believe the location is unneeded. Otherwise, we will register the location to the log.

The problem of cache replacement is distinct from hardware cache with multi-sets. Since the performance cost of the algorithm software such as LRU is high, we design the Log Cache as a direct mapped cache. Therefore, when failing, we directly replace the original cache line. Although this may cause some trashing and other problems, we do not consider these because such scenarios are rare. Also, the replacement and indexing processes may involve consistency problems due to multithreading, which may lead to the problem that improving the redundant rate. But considering the Log Cache is only to reduce the redundant locations quickly, some exceptional cases are not considered.

\subsection{Log Compression}
In the Quantitative Analysis section of Section 3, we have shown if we can reduce the redundant rate and don't increase the time to verifying the redundancy pointers, we can further reduce the performance overhead. Therefore, we propose the model of Figure 8(c). We find that pointers sharing the same ID tend to be stored in adjacent locations, as shown in Figure 8(a). It is a common scenario for C/C++, if programmers need dynamically allocate two-dimensional array. Therefore, one location can represent a contiguous space. We call this mechanism Log Compression. When the Log Compression is used, (1) the location stored in the log metadata will not only represent its location, it may represent 256 or 1024 or more connected locations. As Figure 8 shows, when the Log Compression is not open, the address locations store in log metadata in Figure 8(b) are the address of \emph{buf[5]} and \emph{buf[8]}, while in (c), the ID is stored, so the memory overhead will decrease. (2) locations in Log Cache also serves the same number of contiguous spaces, then the position in the contiguous area is considered to have been stored by returning hit. As the Log Cache also store the object buf\_ID, the procedure 2 in (c) will hit, but in (b), it will miss, because of the same buf\_ID sharing with the address of \emph{buf[5]} and \emph{buf[8]}. Therefore, the redundant rate will further decrease without increasing the verifying time, leading to imporve the performance greatly. (3) When Verify Object(the module used to verify whether there is dangling pointer referring to the object), it is also necessary to determine whether all positions in the entire contiguous space may store a dangling pointer. 

Figure 8(b) and (c) shows two extreme configurations of Log Compression. Figure 8(b) is no Log Compression, while Figure 8(c) is the highest level of Log Compression that using the ID represents the whole object. These two models are very good at showing very poor and good spatial locality, corresponding to computation-intensive and memory-intensive programs. During compiling, the compiler can decide which model to use by static analysis of program characteristics.

\subsection{Delayed Free}
Delayed free is used in dangling pointer elimination tools such as GC. With the pointer graph maintained by DangKiller, we can significantly reduce the searching overhead and free memory objects at more precise timing. As shown in Figure 5, the delayed free module consists of 3 components, Periodical\_Free, Verify Object, and Free\_List.

The Verify Object determines whether a memory object is referenced by a pointer. If such pointer exists, the Verify Object module informs the Delayed\_Free that this object should not be freed. The Verify Object also decides whether an object locates on the heap, if so, the Delayed\_Free will insert that object into the Free\_List.

The Free\_List is a multi-level linked list, as shown in Figure 9. Objects are inserted into Free\_List when they are freed. Upon inserting, Free\_List checks the pointer passed to free() function. If the pointer is located on the stack or refers a global variable, the object is inserted into the ref\_period\_free\_list, waiting for periodically free. If the pointer is stored in another object (objA), the object is inserted into the ref\_heap\_free\_list, as shown in Figure 9, and the root of the head is the objA. Only when the objA is freed, the free\_list of objA will be freed.

\begin{figure}[ht]
  \includegraphics[width=0.48\textwidth]{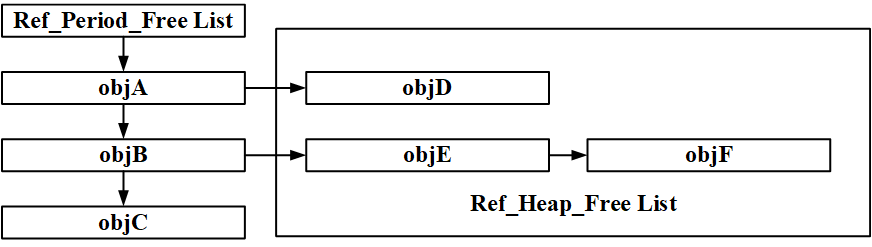}
  \caption[{aboveskip=0.5em, belowskip=0em}]{Free\_List structure}
\end{figure}

The Periodical\_Free periodically decides which objects can be released. Unlike GC, DangKiller does not make this check using multithreading. Instead, we use the length of ref\_period\_free\_list to determine the check timing. if the length is bigger than the threshold, the Periodical\_Free will be executed. The threshold can be user-specified. And a default threshold is 1000. All the evaluation in this paper is based on the default threshold.

\section{Implementation}
We implemented a prototype of Dangkiller on an ARM64 Linux system. DangKiller is implemented in LLVM 3.8 as a compile-time instrumentation pass at the IR level, tracking heap, address-taken stack, and global objects. The auxiliary functions such as minfat\_malloc and pointer\_register are grouped in a separated C file. During instrumentation, the pass will insert calls to these functions into the program IR. By utilizing the Link-Time Optimization feature of LLVM, we do not change the ordinary program build process.

\textbf{Compiler support.}We implemented the pointer tracker, MinFat Pointer memory allocation and free components instrumentation work on the LLVM compiler framework. We use link-time optimizations (LTO) and invoke our pass through the LLVM gold plug-in by the GNU gold linker to run on the LLVM bitcode of the full program.  We use the GNU gold linker instead of the traditional BFD linker and specifying a linker flag. Those features allow our system to be easily integrated into standard build systems for C and C++ programs by only passing compiler flags to enable LTO and invoke the DangKiller pass.

\textbf{Run-time support.}The complete implementation of the run-time functions consists of 1761 LoC, and the libc wrappers contain 5269 LOC. The full implementation of the run-time functions includes registering pointer address to log-structured functions, shared-log structure metadata functions, wrapper functions, Log Cache functions, and delayed free functions.

\section{Optimization}
In the 5.2 section object-metadata address table, we mentioned that the mapping between pointer-to-object needs to combine direct mapped and hash algorithm. However, how to select the boundary between two parts, that is, how to choose the N is especially critical. If N is too large, the memory overhead will rise sharply. If N is too small, it will lead to a large number of hash conflicts and bring more performance overhead. Therefore, it is necessary to select the optimal N based on the experimental data and the characteristics of the program. Also, the same problem occurs in the Log Cache. The size of the cache will affect the miss rate of Log Cache. The larger cache will increase the memory overhead of Log Cache but reduce the performance overhead, and the memory overhead of log metadata structure. However, since DangKiller is a dangling pointer elimination system, it needs more memory than the usual program because a large number of objects cannot be released. Therefore, slightly increasing the capacity of the Log Cache and the N does not raise a lot of memory overhead. So when we design Log Cache and N, we try to increase its capacity for performance. According to the experimental analysis shown in later chapters, we finally choose N to be 32, and the logical size of Log Cache is $2^{28}$.

In addition, whether the Log Compression is on or not is critical to DangKiller. However, different program has different characteristics, so we use the PCA and cluster analysis \cite{RN160} to analysis the characteristics of different programs. And then we use the characteristics to control Dangkiller to select a better strategy when eliminating dangling pointers. And the analysis and the final strategy is shown at \$8.2.

\section{Evaluation}
We evaluate DangKiller and DangSan \cite{RN31} on ARM64 system with SPEC CPU2006. Bogo \cite{RN56,RN34} is not available because it relies on Intel MPX \cite{RN29,RN44} which is not supported by ARM64 architecture. We did not include CRCount \cite{RN54} because it is a closed source software. Additionally, we evaluate DangKiller on SPEC CPU2017. The setup is a AWS 2.3GHz 16 core ARMv8 CPU with 32GB RAM. The performance numbers are the average of 3 runs.

\subsection{Effectiveness}
To test whether Dangkiller can effectively eliminate use-after-free vulnerabilities in real-world applications. We use NIST/Juliet \cite{RN32} benchmarks to verify the effectiveness of our framework. We empirically inspected DangKiller’s elimination capability for 32 cases from NIST/Juliet (CWE416, Use After Free and Double Free) like Bogo \cite{RN53}. Dangkiller soundly eliminated them all.

We benchmarked web servers using Apachebench \cite{RN30,RN2} with the following settings: 64 concurrent connections in the client, 100000 requests issued, and one worker process in the server. We have chosen this configuration to ensure a large amount of concurrency in the webserver process to stress DangKiller’s ability to run the real application. We transfer a tiny file (200 bytes) locally to reduce I/O to a minimum and stress the CPU to provide a conservative estimate of the overhead incurred by DangKiller. With Apache, we achieve 5976 requests per second on the baseline and 5531 when using DangSan. This corresponds with a slowdown of 7\%.

\subsection{Sensitivity Study}
In this part, we study the relationship between Log Compression patterns and program characteristics. Figure 10 shows the performance overhead when applying different Log Compression to  SPEC CPU2006 benchmarks. We can tell that no Log Compression works well for libquantum, bzip2 and astar, while Log Compression works well for mcf, milc, gobmk, and omnetpp.

\begin{figure}[ht]
  \includegraphics[width=0.48\textwidth]{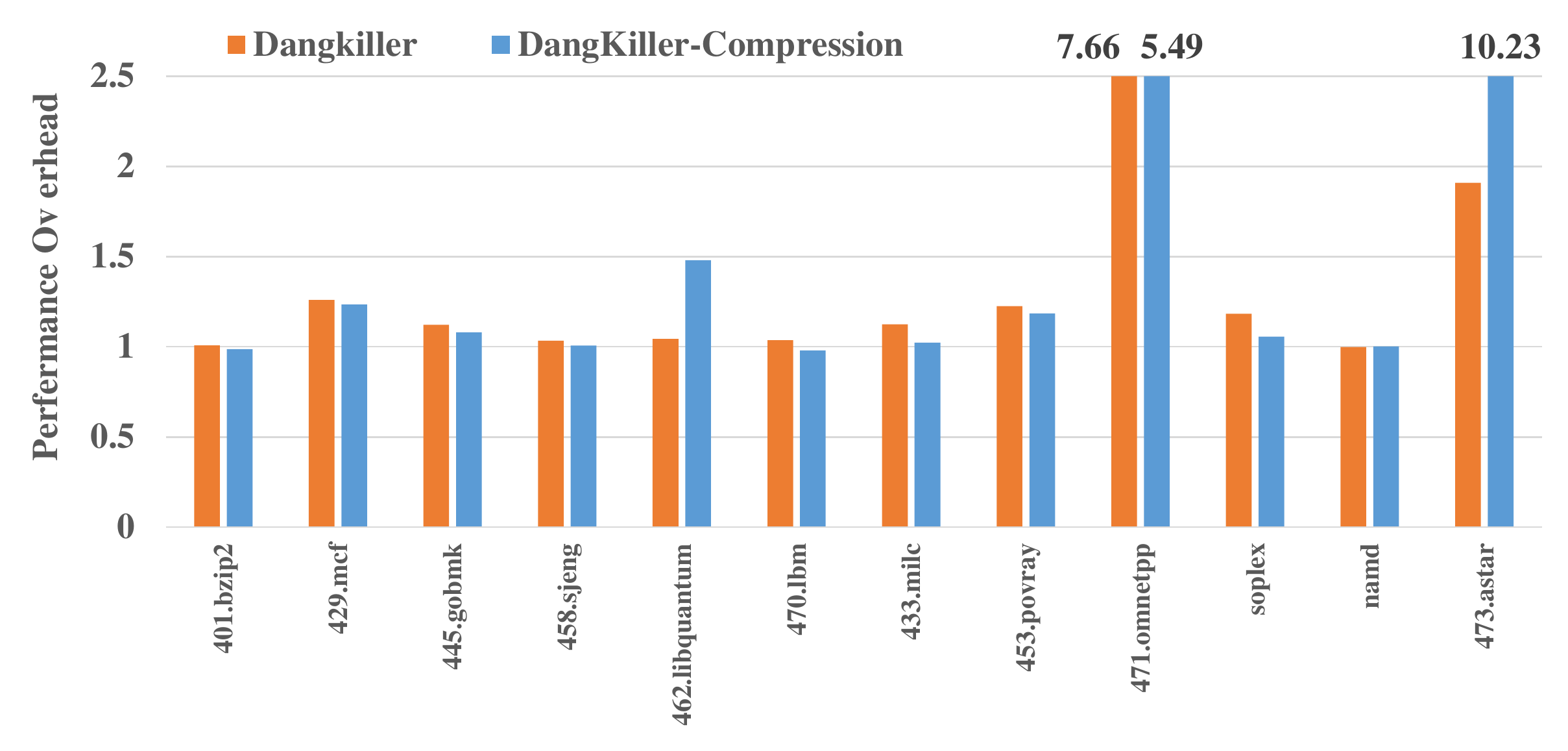}
  \caption[{aboveskip=0.5em, belowskip=0em}]{Compare the performance overhead of with or no Log Compression}
\end{figure}

This is because libquantum, bzip2, and astar are computation-intensive programs which just use the specific locations. On the other hand, memory-intensive programs like mcf and milc usually reference the same object's different offset many times. And some benchmarks like sjeng and lbm which are also computation-intensive benchmarks and have nearly no store operations and object allocations often are not affected by this and have lower performance and memory overhead. Table 1 also demonstrates the same result. From Table 1, we can see when Log Compression is on, the dup rates of mcf and milc are always nearly 95\% . Thinking about the number of  tremendous pointers, we truly reduce a lot of memory and performance overhead. 

\cite{RN160} shows the result of the cluster by using the data of the sampling and profiling on SPEC CPU2006 benchmarks. When the cluster is 4, the mcf, gobmk, and omnetpp are in the same cluster using the Log Compression, while sjeng, astar, and bzip2 are suitable for no Log Compression. The Compiler could decide whether Log Compression or not, according to this information. As for memory overhead, it is mainly decided by the number of pointers being tracked, which is consistent with the characteristics of performance overhead.

\subsection{Performance Overhead}
We run SPEC CPU2006 and CPU2017 with DangKiller to evaluate the performance overhead. The results are shown in Figure 11 and 12. Some of the benchmarks will malfunction when compiled with DangKiller, because the tagged pointer scheme we used to track the relationships between objects and pointers has inherent compatibility issues. For example, benchmarks, like gcc, actually utilize the high bits of pointers, so our tagging process will break the original program.  

\begin{figure}[ht]
  \includegraphics[width=0.48\textwidth]{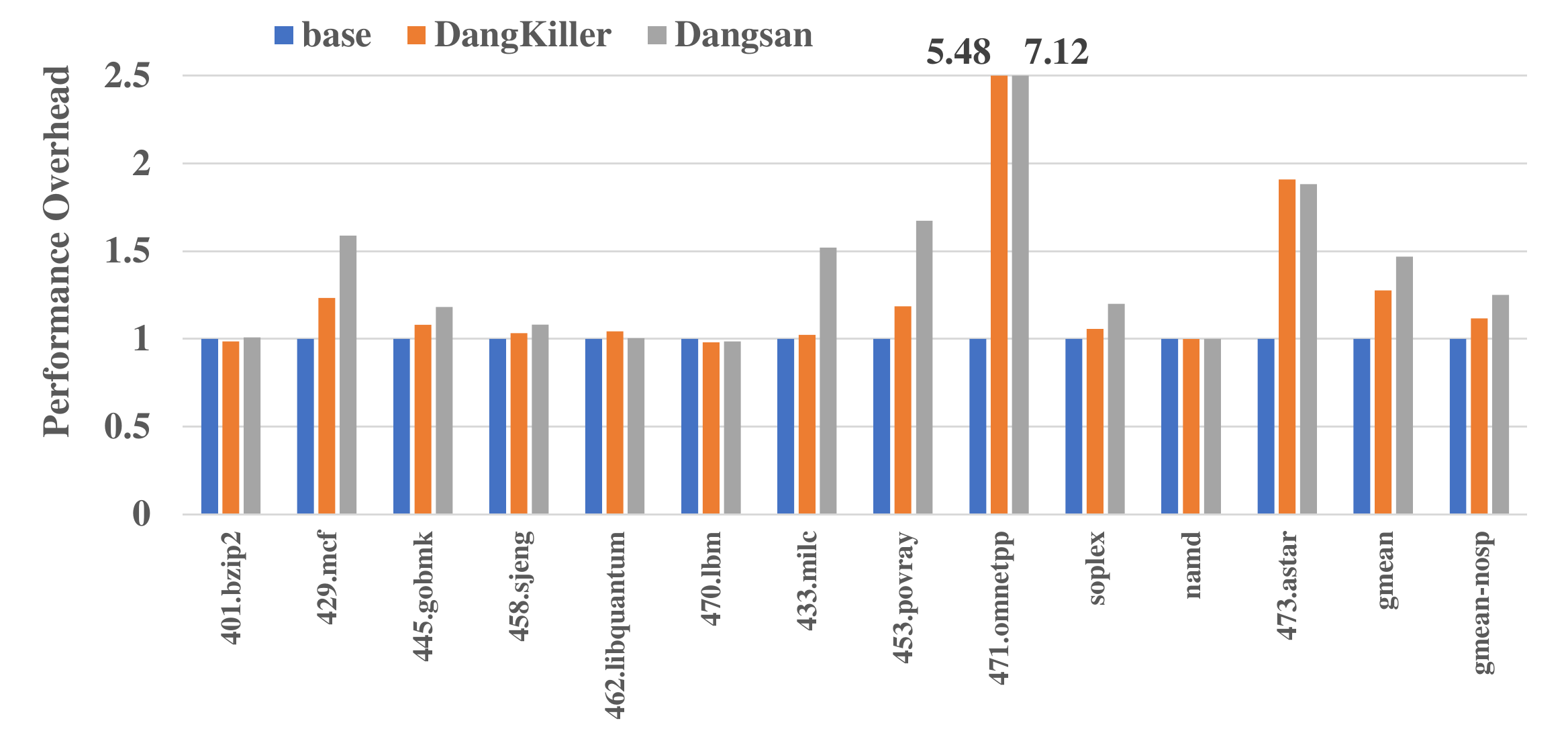}
  \caption[{aboveskip=0.5em, belowskip=0em}]{Performance Overhead on SPEC CPU2006}
\end{figure}
\begin{figure}[ht]
  \includegraphics[width=0.48\textwidth]{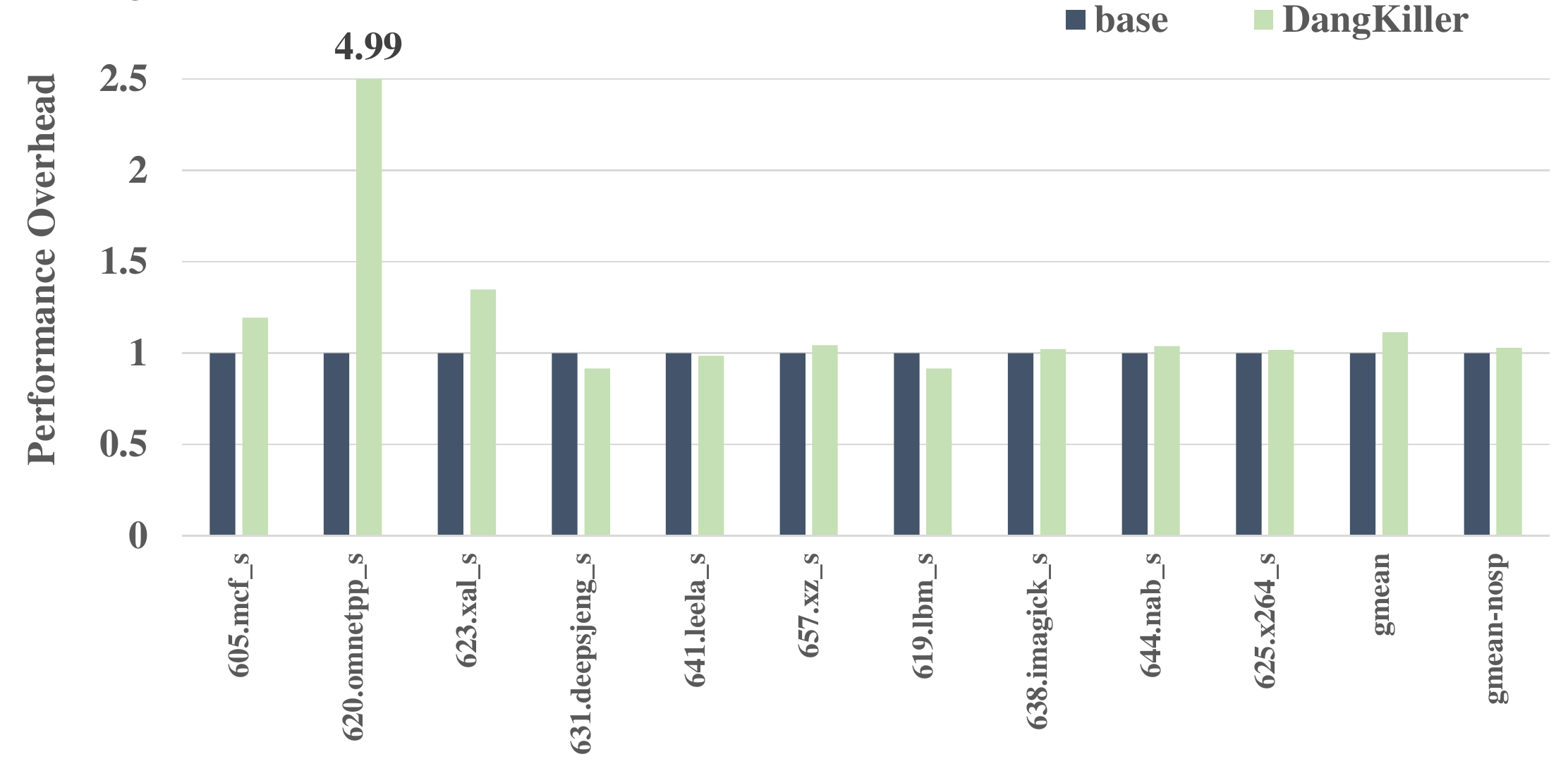}
  \caption[{aboveskip=0.5em, belowskip=0em}]{Performance Overhead on SPEC CPU2017}
\end{figure}

The performance results for the C and C++ programs of the SPEC CPU2006 benchmark suite on the ARM64 system are presented as Figure 11. Compared with the results reported by DangSan, some benchmarks such as bzip2, lbm, and sjeng, DangKiller's performance is almost the same as Dangsan. On these benchmarks, both of the tools have a small performance overhead, and these data can not fully show the difference in performance overhead between the two methods. However, on the other benchmarks, especially benchmarks like mcf, milc, astar, omnetpp, etc., DangKiller can significantly reduce the performance overhead. At the same time,  DangKiller has better performance overhead than DangSan in each benchmark shown in Figure 11, especially for many memory-intensive programs. Similar to SPEC CPU2006 (Dangsan can't run most of the SPEC CPU2017 benchmarks), except for several benchmarks with compatibility issues, SPEC CPU2017's evaluation data shows the same features as shown in the Figure 12, but our DangKiller has a better performance in the SPEC CPU2017, which is closer to the current real situation. Except for omnetpp, almost all benchmarks have only about 5\% performance overhead, and some benchmarks, such as deepsjeng and lbm, even have about 5\% performance improvement. The performance improvement mainly attributes to that some objects won't really execute the libc-free function until the end of program, because there are still some dangling pointers reference to them.  

DangKiller has an average of 25\% and 11\% performance overhead, while DangSan has nearly 50\% performance overhead on ARM64 systems when running the same benchmarks. The omnetpp benchmark has an extremely high performance overhead, which is because of the large memory access operations. All the pointer tracking schemes will inevitably encounter a similar slowdown on this benchmark. Except for omnetpp, DangKiller has only 11\% and 3\% performance loss. Above all, we can get the conclusion that DangKiller is the best dangling pointer elimination system, although there are a few benchmarks limited by Minfat Pointer and ARM64 system compatibility issues.

Additionally, for Bogo using MPX technology, the average performance loss is about 34\%, and in terms of average performance, DangKiller is also better than Bogo. For CRCount, when using an implicit invalid dangling pointer on an x86 platform. However, we noticed a performance difference between an ARM64 system and an x86 system. But considering the difference between ARM64 system and x86 system, for example, DangSan has about 40\% overhead on x86 system, but on ARM64, there is a 50\% overhead on the system. So CRCount may have a small increase in overhead on the ARM64 system, too. On the other hand, CRCount can only partially eliminate dangling pointer, and we have the confidence to believe our DangKiller still has a significant advantage.

\subsection{Memory Overhead}
In the DangKiller, except for data structures, undeleted objects are the major factor that potentially consumes substantial memory. To determine the impact of our system on memory usage, we have measured the mean resident set size (RSS) while running the SPEC CPU2006 benchmark suite.

\begin{figure}[ht]
  \includegraphics[width=0.48\textwidth]{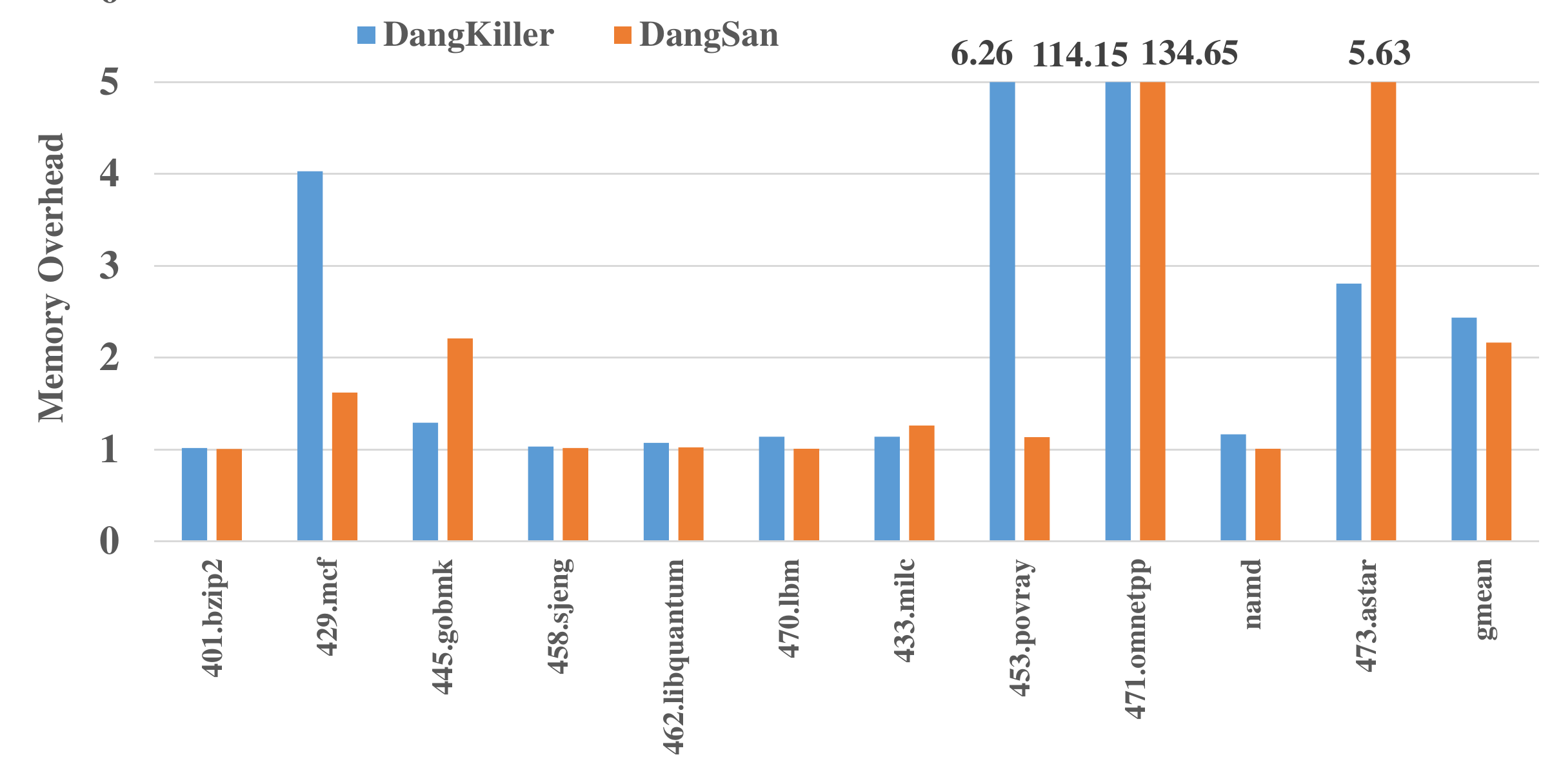}
  \caption[{aboveskip=0.5em, belowskip=0em}]{Memory Overhead on SPEC CPU2006}
\end{figure}

Figure 13 shows the memory overhead of our DangSan and DangKiller for SPEC CPU2006 benchmarks. Our geometric mean of all benchmarks is 143.5\%, which is higher than 126.4\% of DangSan and other existing works. However, Dangsan's memory overhead comes only from its complex metadata structure, while DangKiller is an elimination technique that delays the free of many objects, resulting in considerable objects accumulation, which leads to more memory overhead. For example, mcf uses the realloc function, rather than detection tools nullifying the dangling pointers, DangKiller waiting for the former objects' free, which has nearly four times memory overhead. In fact, such memory overhead is unavoidable. Besides, DangKiller needs to allocate and align the object powers of two, which makes the memory overhead a bit higher. For Bogo, which reuses spatial metadata, and CRCount, which only retains reference counts, Dangkiller does have a higher memory overhead. In fact, except for omnetpp and astar, most programs only have an average of 89\% of memory overhead, but considering that in the environment of memory free, it is acceptable to increase memory overhead in pursuit of some performance improvements. Of course, if you are applying to a memory-constrained environment, you can refer to the method of optimizing Log Cache and shared hash address space in Section 7 to reduce the memory overhead to a suitable way.

\begin{table}[ht]
  \centering
  \begin{tabular}{|c|c|c|c|c|c|c|c|}
    \hline
    
    \multirow{2}{*} {\small{benchmark}} &  \multirow{2}{*}{\#obj} &  \multirow{2}{*}{\#ptrs} & \multicolumn{2}{|c|}{\#dup } &  \multirow{2}{*}{\# free} \\
                   ~          &     ~   &    ~   & \multicolumn{2}{|c|}{\ DSan/DKiller }  &   ~ \\
   
    \hline
    bzip2 & 258 & 2200k & 84.9\% & 97.7\% & 144\\
    \hline
    mcf & 20 & 7658m &  99.3\% &  99.5\% & 7  \\
    \hline
    gobmk & 622k & 607m & 98.3\% & 99.7\% & 657k \\
    \hline
    sjeng & 20 & 4 &  0\% & 0\% & 0 \\
    \hline
    \small{libquantum} & 164 & 130 & 23.1\%  & 70\% & 179\\
    \hline 
    lbm & 19 & 6004 & 50\% & 99.9\% & 2 \\
    \hline
    milc & 6530 & 2585m & 61.9\% & 95.3\% & 6468 \\
    \hline
    povray & 2427k & 4679m & 95.3\% & 97.1\%  & 2461k \\
    \hline
    omnetpp & 267m & \small{13099m} & 70.3\% & 80.4\% & 267m \\
    \hline
    namd & 1339 & 2970k & 62.8\% & 84.4\% &  1304\\
    \hline
    astar & 4800k & 1235m & 89.9\% & 89.9\% & 4799k \\
    \hline
   
  \end{tabular}
  \caption{Statistics for SPEC CPU2006.}
  \label{table:formatting}
\end{table}
\subsection{Coverage and Statistics}
Table 1 shows the number of memory objects (\#obj alloc) and pointers (\#ptrs) we tracked, the number of duplicate pointers (\#dup) in DangKiller and Dangsan, and how many times DangKiller tried to free objects (\#free). As shown in the \#dup column, we can see that DangKiller identified much more duplicate pointers than DangSan, especially in some memory-intensive benchmarks like milc and omnetpp. Therefore, the Log Compression mechanism can significantly reduce the redundancy of log metadata, thus improving the performance.

Due to terminating the Verify Object process immediately after finding dangling pointer, we did not count the number of dangling pointers when an object is released. Instead, Table 1 shows the total times when the delayed free module tried to free an object. We found that most of the program in SPEC CPU2006 allocated many objects without freeing them, especially in some benchmarks like bizp2 and sjeng. Therefore, although the times of free are less than the times of allocation, there were actually a lot of delayed frees. Besides, if an object is not freed explicitly by the user, we do not free that object.
\section{Limitation}
While our DangKiller should be able to eliminate nearly all dangling pointer vulnerabilities in practical settings efficiently, it has a few fundamental limitations. We discuss these limitations in this section.

There are several limitations faced by every approach based on pointer tracking \cite{RN23,RN54,RN52,RN28}. The DangKiller cannot track metadata for uninstrumented shared libraries. Also, we only track pointers that are explicitly stored in llvm IR bytecode, which means that there is one case in which we do not track: pointers are spilled onto the stack by a function prologue and will be restored to a register when the function returns to its caller because spilling is triggered by register allocation which usually done by CodeGen and MC layer in llvm. Moreover, Dangkiller doesn't track pointers copied in a type-unsafe way. For example, pointers cast to integers and pointers copied by the memcpy function. Though Dangkiller has wrappers on std-lib and LTO passes which can track all the potential type-unsafe pointers, we think there would be a minimal risk of false negatives because such a scenario is rare. Also, Dangkiller may be prone to false positives due to type accuracy issues. However, due to the use of MinFat Pointer, a pointer with tag information makes this type of accuracy error less likely than existing work.

Last, although MinFat Pointer is applied to the ARM64 platform with Address Tagging, some programs that perform particular operations on pointers, such as gcc that manipulate high bits of pointers may still cause some compatibility problems. All the tagged pointer schemes share the same compatibility issue.
\section{Future work}
Like many existing elimination works, our delayed free technique can be implemented with multithreading to reduce its performance overhead, e.g., periodically using other threads to detect whether a object in the free\_list can be released. We did not do this because we wanted to implement a lightweight, dangling pointer elimination tool that can even be suitable for IoT devices. Implementing a multithreaded scheme on such devices would consume too much of the already limited computational resources, and may exceed the power limitation. However, for desktop and server level applications, if there are additional requirements for performance, we can also support multithreading features in the future.

Some latest works consider memcpy, which can be used to copy pointers. Since our work has wrapper for memcpy, we can track the memcpy in the wrapper function. We believe that can be developed in the futuer.


Finally, we do not support memory objects collection like GC. However, combining with the  multithreading and our efficient pointer graph management, we believe that we can design a lightweight garbage collection for C/C++ in the near future.
\section{Conclusion}
We presented the DangKiller, a novel, and lightweight dangling pointer elimination system with low memory and performance overhead. The DangKiller allows a wide range of C/C++ applications to tolerate dangling pointer vulnerabilities. We generate unique IDs shared by pointers and their intended referent with a tagged pointer system, MinFat Pointer. The DangKiller utilizes these IDs to build a high-performance pointer tracking model and support the delayed free technique. The free of an object is delayed until its references disappear during program execution, thus preventing the generation of dangling pointers. We also proposed an efficient Log Cache and a Log Compression mechanism, which significantly reduced the metadata searching overheads and the redundancy in dangling pointer candidates. Attaining virtually more safety guarantee than other pointer invalidation solutions, the DangKiller incurs 11\% and 3\% performance overheads respectively on SPEC CPU2006 and CPU2017 benchmarks except for some special memory-intensive cases. To be concluded, DangKiller is more efficient than existing dangling pointer detecting systems and can support multithreading by removing complex pointer-to-object mapping and reducing the redundancy of locations. To foster further research in the field, we have made the source code of our DangKiller prototype available as open-source upon acceptance.

\bibliographystyle{plain}
\bibliography{references}

\begin{thebibliography}{10}

\bibitem{RN29}
2011 cwe/sans top 25 most dangerous software errors.
\newblock \url{http://cwe.mitre.org/top25/}.

\bibitem{RN30}
ab - apache http server benchmarking tool.
\newblock \url{https://httpd.apache.org/docs/2.4/programs/ab.html}.

\bibitem{RN51}
Addresssanitizer.
\newblock \url{https://github.com/google/sanitizers/wiki}.

\bibitem{RN165}
addresssanitizer design documentation.
\newblock
  \url{http://clang.llvm.org/docs/HardwareAssistedAddressSanitizerDesign.html}.

\bibitem{RN2}
Apache.
\newblock \url{https://www.apache.org/}.

\bibitem{RN167}
Armv8-a address tagging documentation.
\newblock
  \url{http://infocenter.arm.com/help/index.jsp?topic=/com.arm.doc.den0024a/ch12s05s01.html}.

\bibitem{RN56}
Bogo open source implementation.
\newblock \url{https://github.com/lzto/bogo/}.

\bibitem{RN31}
Dangsan open source implementation.
\newblock \url{https://github.com/vusec/dangsan}.

\bibitem{RN32}
Nist software assurance reference dataset project.
\newblock \url{https://samate.nist.gov/SARD}.

\bibitem{RN34}
Struct bound narrowing.
\newblock \url{https://gcc.gnu.org/wiki/Intel MPX support in the GCC
  compiler#Narrowing}.

\bibitem{RN7}
Periklis Akritidis.
\newblock Cling: A memory allocator to mitigate dangling pointers.
\newblock In {\em Usenix Conference on Security}.

\bibitem{RN8}
Periklis Akritidis, Manuel Costa, Miguel Castro, and Steven Hand.
\newblock Baggy bounds checking: An efficient and backwards-compatible defense
  against out-of-bounds errors.
\newblock {\em Proc Usenix Ssym}, pages 51--66, 2010.

\bibitem{RN57}
Sergei Arnautov, Bohdan Trach, Franz Gregor, Thomas Knauth, Andre Martin,
  Christian Priebe, Joshua Lind, Divya Muthukumaran, Dan O'Keeffe, and Mark~L
  Stillwell.
\newblock {SCONE}: Secure linux containers with intel {SGX}.
\newblock In {\em 12th {USENIX} Symposium on Operating Systems Design and
  Implementation ({OSDI} 16)}, pages 689--703, 2016.

\bibitem{RN9}
Emery~D. Berger and Benjamin~G. Zorn.
\newblock Diehard:probabilistic memory safety for unsafe languages.
\newblock {\em Acm Sigplan Notices}, 41(6):158--168, 2006.

\bibitem{RN169}
Tyler Bletsch, Xuxian Jiang, Vince~W Freeh, and Zhenkai Liang.
\newblock Jump-oriented programming: a new class of code-reuse attack.
\newblock In {\em Proceedings of the 6th ACM Symposium on Information, Computer
  and Communications Security}, pages 30--40. ACM, 2011.

\bibitem{RN11}
H.~Boehm.
\newblock A garbage collector for c and c++.
\newblock \url{http://www.hboehm.info/gc/}.

\bibitem{RN164}
Jeremy Brown, JP~Grossman, Andrew Huang, and Thomas~F Knight~Jr.
\newblock A capability representation with embedded address and nearly-exact
  object bounds.
\newblock {\em Project Aries Technical Memo 5, Tech. Rep.}, 2000.

\bibitem{RN184}
Nicholas~P Carter, Stephen~W Keckler, and William~J Dally.
\newblock Hardware support for fast capability-based addressing.
\newblock In {\em ACM SIGPLAN Notices}, volume~29, pages 319--327. ACM, 1994.

\bibitem{RN183}
Nicholas~P Carter, Stephen~W Keckler, and William~J Dally.
\newblock Memory system including guarded pointers, December~1 1998.
\newblock US Patent 5,845,331.

\bibitem{RN33}
Stephen Checkoway, Lucas Davi, Alexandra Dmitrienko, Ahmad-Reza Sadeghi, Hovav
  Shacham, and Marcel Winandy.
\newblock Return-oriented programming without returns.
\newblock In {\em Proceedings of the 17th ACM conference on Computer and
  communications security}, pages 559--572. ACM, 2010.

\bibitem{RN15}
Jim Chow, Ben Pfaff, Tal Garfinkel, and Mendel Rosenblum.
\newblock Shredding your garbage: Reducing data lifetime through secure
  deallocation.
\newblock In {\em USENIX Security Symposium}, pages 22--22, 2005.

\bibitem{RN128}
Thurston~HY Dang, Petros Maniatis, and David Wagner.
\newblock Oscar: A practical page-permissions-based scheme for thwarting
  dangling pointers.
\newblock In {\em 26th {USENIX} Security Symposium ({USENIX} Security 17)},
  pages 815--832, 2017.

\bibitem{RN35}
Dinakar Dhurjati and Vikram Adve.
\newblock Backwards-compatible array bounds checking for c with very low
  overhead.
\newblock In {\em Proceedings of the 28th international conference on Software
  engineering}, pages 162--171. ACM, 2006.

\bibitem{RN36}
Dinakar Dhurjati and Vikram Adve.
\newblock Efficiently detecting all dangling pointer uses in production
  servers.
\newblock In {\em International Conference on Dependable Systems and Networks
  (DSN'06)}, pages 269--280. IEEE, 2006.

\bibitem{RN37}
Dinakar Dhurjati, Sumant Kowshik, Vikram Adve, and Chris Lattner.
\newblock Memory safety without garbage collection for embedded applications.
\newblock {\em ACM Transactions on Embedded Computing Systems (TECS)},
  4(1):73--111, 2005.

\bibitem{RN162}
Gregory~J Duck and Roland~HC Yap.
\newblock Heap bounds protection with low fat pointers.
\newblock In {\em Proceedings of the 25th International Conference on Compiler
  Construction}, pages 132--142. ACM, 2016.

\bibitem{RN38}
Gregory~J Duck and Roland~HC Yap.
\newblock Effectivesan: type and memory error detection using dynamically typed
  c/c++.
\newblock In {\em ACM SIGPLAN Notices}, volume~53, pages 181--195. ACM, 2018.

\bibitem{RN161}
Gregory~J Duck, Roland~HC Yap, and Lorenzo Cavallaro.
\newblock Stack bounds protection with low fat pointers.
\newblock In {\em The Network and Distributed System Security Symposium
  (NDSS)}, 2017.

\bibitem{RN16}
Josselin Feist, Laurent Mounier, and Marie-Laure Potet.
\newblock Statically detecting use after free on binary code.
\newblock {\em Journal of Computer Virology and Hacking Techniques},
  10(3):211--217, 2014.

\bibitem{RN181}
Sanjay Ghemawat and Paul Menage.
\newblock Tcmalloc: Thread-caching malloc, 2009.

\bibitem{RN17}
Cristiano Giuffrida, Anton Kuijsten, and Andrew~S Tanenbaum.
\newblock Enhanced operating system security through efficient and fine-grained
  address space randomization.
\newblock In {\em Presented as part of the 21st {USENIX} Security Symposium
  ({USENIX} Security 12)}, pages 475--490, 2012.

\bibitem{RN182}
Wolfram Gloger.
\newblock Ptmalloc.
\newblock {\em Consult{\'e} sur http://www. malloc. de/en}, 2006.

\bibitem{RN18}
Ben Gras, Kaveh Razavi, Erik Bosman, Herbert Bos, and Cristiano Giuffrida.
\newblock Aslr on the line: Practical cache attacks on the mmu.
\newblock In {\em The Network and Distributed System Security Symposium
  (NDSS)}, volume~17, page~26, 2017.

\bibitem{RN19}
Istvan Haller, Yuseok Jeon, Hui Peng, Mathias Payer, Cristiano Giuffrida,
  Herbert Bos, and Erik Van Der~Kouwe.
\newblock Typesan: Practical type confusion detection.
\newblock In {\em Proceedings of the 2016 ACM SIGSAC Conference on Computer and
  Communications Security}, pages 517--528. ACM, 2016.

\bibitem{RN174}
Hong Hu, Shweta Shinde, Sendroiu Adrian, Zheng~Leong Chua, Prateek Saxena, and
  Zhenkai Liang.
\newblock Data-oriented programming: On the expressiveness of non-control data
  attacks.
\newblock In {\em 2016 IEEE Symposium on Security and Privacy (SP)}, pages
  969--986. IEEE, 2016.

\bibitem{RN21}
Yeongjin Jang, Sangho Lee, and Taesoo Kim.
\newblock Breaking kernel address space layout randomization with intel tsx.
\newblock In {\em Proceedings of the 2016 ACM SIGSAC Conference on Computer and
  Communications Security}, pages 380--392. ACM, 2016.

\bibitem{RN22}
Koen Koning, Xi~Chen, Herbert Bos, Cristiano Giuffrida, and Elias
  Athanasopoulos.
\newblock No need to hide: Protecting safe regions on commodity hardware.
\newblock In {\em Proceedings of the Twelfth European Conference on Computer
  Systems}, pages 437--452. ACM, 2017.

\bibitem{RN40}
Taddeus Kroes, Koen Koning, Erik van~der Kouwe, Herbert Bos, and Cristiano
  Giuffrida.
\newblock Delta pointers: Buffer overflow checks without the checks.
\newblock In {\em Proceedings of the Thirteenth EuroSys Conference}, page~22.
  ACM, 2018.

\bibitem{RN163}
Albert Kwon, Udit Dhawan, Jonathan~M Smith, Thomas~F Knight~Jr, and Andre
  DeHon.
\newblock Low-fat pointers: compact encoding and efficient gate-level
  implementation of fat pointers for spatial safety and capability-based
  security.
\newblock In {\em Proceedings of the 2013 ACM SIGSAC conference on Computer \&
  communications security}, pages 721--732. ACM, 2013.

\bibitem{RN175}
Chris Lattner and Vikram Adve.
\newblock Llvm: A compilation framework for lifelong program analysis \&
  transformation.
\newblock In {\em Proceedings of the international symposium on Code generation
  and optimization: feedback-directed and runtime optimization}, page~75. IEEE
  Computer Society, 2004.

\bibitem{RN23}
Byoungyoung Lee, Chengyu Song, Yeongjin Jang, Tielei Wang, Taesoo Kim, Long Lu,
  and Wenke Lee.
\newblock Preventing use-after-free with dangling pointers nullification.
\newblock In {\em The Network and Distributed System Security Symposium
  (NDSS)}, 2015.

\bibitem{RN179}
Daiping Liu, Mingwei Zhang, and Haining Wang.
\newblock A robust and efficient defense against use-after-free exploits via
  concurrent pointer sweeping.
\newblock In {\em Proceedings of the 2018 ACM SIGSAC Conference on Computer and
  Communications Security}, pages 1635--1648. ACM, 2018.

\bibitem{RN171}
Matt Miller.
\newblock Trends and challenges in the vulnerability mitigation landscape.
\newblock In {\em 13th {USENIX} Workshop on Offensive Technologies ({WOOT}
  19)}, Santa Clara, CA, August 2019. {USENIX} Association.

\bibitem{RN42}
Santosh Nagarakatte, Milo~MK Martin, and Steve Zdancewic.
\newblock Watchdog: Hardware for safe and secure manual memory management and
  full memory safety.
\newblock In {\em 2012 39th Annual International Symposium on Computer
  Architecture (ISCA)}, pages 189--200. IEEE, 2012.

\bibitem{RN43}
Santosh Nagarakatte, Jianzhou Zhao, Milo~MK Martin, and Steve Zdancewic.
\newblock Softbound: Highly compatible and complete spatial memory safety for
  c.
\newblock {\em ACM Sigplan Notices}, 44(6):245--258, 2009.

\bibitem{RN24}
Santosh Nagarakatte, Jianzhou Zhao, Milo~MK Martin, and Steve Zdancewic.
\newblock Cets: compiler enforced temporal safety for c.
\newblock {\em ACM Sigplan Notices}, 45(8):31--40, 2010.

\bibitem{RN25}
Gene Novark and Emery~D Berger.
\newblock Dieharder: securing the heap.
\newblock In {\em Proceedings of the 17th ACM conference on Computer and
  communications security}, pages 573--584. ACM, 2010.

\bibitem{RN44}
Oleksii Oleksenko, Dmitrii Kuvaiskii, Pramod Bhatotia, Pascal Felber, and
  Christof Fetzer.
\newblock Intel mpx explained: An empirical study of intel mpx and
  software-based bounds checking approaches.
\newblock {\em arXiv preprint arXiv:1702.00719}, 2017.

\bibitem{RN160}
Aashish Phansalkar, Ajay Joshi, and Lizy~K John.
\newblock Analysis of redundancy and application balance in the spec cpu2006
  benchmark suite.
\newblock {\em ACM SIGARCH Computer Architecture News}, 35(2):412--423, 2007.

\bibitem{RN177}
Martin Rinard.
\newblock Acceptability-oriented computing.
\newblock {\em Acm sigplan notices}, 38(12):57--75, 2003.

\bibitem{RN176}
Martin~C Rinard, Cristian Cadar, Daniel Dumitran, Daniel~M Roy, Tudor Leu, and
  William~S Beebee.
\newblock Enhancing server availability and security through failure-oblivious
  computing.
\newblock In {\em OSDI}, volume~4, pages 21--21, 2004.

\bibitem{RN26}
Konstantin Serebryany, Derek Bruening, Alexander Potapenko, and Dmitriy Vyukov.
\newblock Addresssanitizer: A fast address sanity checker.
\newblock In {\em Presented as part of the 2012 {USENIX} Annual Technical
  Conference ({USENIX}{ATC} 12)}, pages 309--318, 2012.

\bibitem{RN166}
Kostya Serebryany, Evgenii Stepanov, Aleksey Shlyapnikov, Vlad Tsyrklevich, and
  Dmitry Vyukov.
\newblock Memory tagging and how it improves c/c++ memory safety.
\newblock {\em arXiv preprint arXiv:1802.09517}, 2018.

\bibitem{RN54}
Jangseop Shin, Donghyun Kwon, Jiwon Seo, Yeongpil Cho, and Yunheung Paek.
\newblock Crcount: Pointer invalidation with reference counting to mitigate
  use-after-free in legacy c/c++.
\newblock In {\em The Network and Distributed System Security Symposium
  (NDSS)}, 2019.

\bibitem{RN170}
Minh Tran, Mark Etheridge, Tyler Bletsch, Xuxian Jiang, Vincent Freeh, and Peng
  Ning.
\newblock On the expressiveness of return-into-libc attacks.
\newblock In {\em International Workshop on Recent Advances in Intrusion
  Detection}, pages 121--141. Springer, 2011.

\bibitem{RN52}
Erik Van Der~Kouwe, Vinod Nigade, and Cristiano Giuffrida.
\newblock Dangsan: Scalable use-after-free detection.
\newblock In {\em Proceedings of the Twelfth European Conference on Computer
  Systems}, pages 405--419. ACM, 2017.

\bibitem{RN49}
Jonathan Woodruff, Robert~NM Watson, David Chisnall, Simon~W Moore, Jonathan
  Anderson, Brooks Davis, Ben Laurie, Peter~G Neumann, Robert Norton, and
  Michael Roe.
\newblock The cheri capability model: Revisiting risc in an age of risk.
\newblock In {\em 2014 ACM/IEEE 41st International Symposium on Computer
  Architecture (ISCA)}, pages 457--468. IEEE, 2014.

\bibitem{RN178}
Hongyan Xia, Jonathan Woodruff, Sam Ainsworth, Nathaniel~W Filardo, Michael
  Roe, Alexander Richardson, Peter Rugg, Peter~G Neumann, Simon~W Moore,
  Robert~NM Watson, et~al.
\newblock Cherivoke: Characterising pointer revocation using cheri capabilities
  for temporal memory safety.
\newblock In {\em Proceedings of the 52nd Annual IEEE/ACM International
  Symposium on Microarchitecture}, pages 545--557. ACM, 2019.

\bibitem{RN180}
Wen Xu, Juanru Li, Junliang Shu, Wenbo Yang, Tianyi Xie, Yuanyuan Zhang, and
  Dawu Gu.
\newblock From collision to exploitation: Unleashing use-after-free
  vulnerabilities in linux kernel.
\newblock In {\em Proceedings of the 22nd ACM SIGSAC Conference on Computer and
  Communications Security}, pages 414--425. ACM, 2015.

\bibitem{RN28}
Yves Younan.
\newblock Freesentry: protecting against use-after-free vulnerabilities due to
  dangling pointers.
\newblock In {\em The Network and Distributed System Security Symposium
  (NDSS)}, 2015.

\bibitem{RN53}
Tong Zhang, Dongyoon Lee, and Changhee Jung.
\newblock Bogo: Buy spatial memory safety, get temporal memory safety (almost)
  free.
\newblock In {\em Proceedings of the Twenty-Fourth International Conference on
  Architectural Support for Programming Languages and Operating Systems}, pages
  631--644. ACM, 2019.

\end{thebibliography}

\end{document}